\documentclass{aa}  

\makeatletter
\renewcommand*\aa@manuscriptname{%
  manuscript no. aa60557-26
  \hspace{\stretch{1}}%
  \copyright ESO \the\year
}
\makeatother
\usepackage{natbib}
\usepackage{graphicx}
\usepackage{txfonts}

\begin{document}

   \title{Exploring if a coronal dimming event can produce coronal hole-like properties}

   \author{Adriana De-Sassi\inst{1}\fnmsep\thanks{Corresponding author: adesassi@ethz.ch}
        \and Louise Harra\inst{1,}\inst{2}
        \and Krzysztof Barczynski\inst{2}
        \and David H. Brooks\inst{3,4,5}
        \and  Lakshmi Pradeep Chitta \inst{8}
        \and Ioannis Kontogiannis\inst{1}
        \and Säm Krucker \inst{10}
        \and Cristina H. Mandrini\inst{6}
        \and Eric Priest \inst{9}
        \and Alphonse C. Sterling \inst{7}
        \and Yingjie Zhu \inst{1}
        }

   \institute{ETH Zurich, Institute for Particle Physics and Astrophysics, Wolfgang-Pauli-Strasse 27, 8092 Zurich, Switzerland
   \and PMOD/WRC, Dorfstrasse 33, 7260 Davos Dorf, Switzerland
   \and Computational Physics Inc., Springfield, VA 22151, USA
   \and University College London, Mullard Space Science Laboratory, Holmbury St. Mary, Dorking, Surrey, RH5 6NT, UK
   \and National Institutes of Natural Sciences, National Astronomical Observatory of Japan, 2-21-1 Osawa, Mitaka, Tokyo, 181-8588, Japan
   \and Instituto de Astronomía y Física del Espacio, IAFE, UBA-CONICET, Pab. IAFE, Ciudad Universitaria, 1428 Buenos Aires, Argentina
   \and NASA/Marshall Space Flight Center, Huntsville, AL 35812, USA
   \and  Max-Planck-Institut für Sonnensystemforschung, 37077 Göttingen, Germany 
   \and School of Mathematics and Statistics, University of St Andrews, St Andrews, Fife KY16 9SS, UK
   \and University of Applied Sciences and Arts Northwestern Switzerland, 5210 Windisch, Switzerland}

   \date{Received 12 April 2026 / Accepted 10 July 10 2026}
 
  \abstract
  {Coronal dimmings, or transient coronal holes, are manifested as a sudden reduction in extreme ultraviolet (EUV) and X-ray emission, often following solar eruptions.}
  {We investigate whether a dimming event can produce coronal hole-like plasma characteristics in the initially quiet Sun.}
  {On 17 and 18 March 2022, Solar Orbiter, Hinode, IRIS, and SDO observed a coronal dimming event that occurred after a quiescent filament eruption. 
  We analysed imaging and spectroscopic data to compare the chromospheric and coronal response prior to and during the coronal dimming. 
  } 
  {The SDO/AIA {193\;\AA} emission intensity in the dimming region was reduced to the same level as the neighbouring coronal hole within 11\,hours. The Doppler velocity measured with \ion{Fe}{XII} (corona) decreased from  $-0.34^{+0.37}_{-0.27}\,\mathrm{km\,s^{-1}}$ towards a predominant upflow of $-3.2^{+0.4}_{-0.6}\,\mathrm{km\,s^{-1}}$. The first ionisation potential (FIP) bias was reduced towards photospheric values. We found an increase in the number of automatically detected EUV brightenings near the dimming boundary in SDO/AIA 193~\AA\,, which could be a sign of magnetic reconnection. In the cooler SDO/AIA 171~{\AA} or Solar Orbiter HRI$_\text{EUV}$ 174 \AA\; channel, we did not observe such an increase. Coronal bright points (CBPs) appeared relatively unaffected by the formation of the dimming. The \ion{Mg}{II} $k_3$ (chromosphere) Doppler velocities were unchanged, except for a small reduction in the dimming upflows in areas with weak magnetic fields ($<20$ G).}
{We find that the dimming only shows partial coronal hole-like properties; specifically, in the coronal emission lines and at temperatures of $>1$ MK. We suggest that this is due to the dimming resulting from plasma depletion only at higher altitudes. Since the CBPs were not significantly impacted by the dimming, we used their magnetic loop heights ($\sim 10$~Mm) as the lower limit to the dimming height. Our findings provide new insights into the atmospheric structure of coronal dimmings. }


   \maketitle
   \nolinenumbers

\section{Introduction}
Coronal dimmings were first observed with coronagraphs as abrupt depletions in white light intensity \citep{hansenAbruptDepletionsInner1974}. Subsequently, they were directly observed on the solar disk as temporary reductions in the X-ray emission \citep[e.g.][]{rustCoronalDisturbancesTheir1983, hudsonCoronalXRayDimming1996, sterlingYohkohSXTObservations1997a} and extreme ultraviolet (EUV) emission \citep[e.g.][]{thompsonSOHOEITObservations1998, zarroSOHOEITObservations1999}.   

Dimming regions are closely associated with coronal mass ejections \citep{bewsherRelationshipEUVDimming2008}. The dimming is thought to be caused by reduced density due to evacuated coronal material \citep{harrison2000spectroscopic}. A possible explanation is that the material escapes along newly opened magnetic field lines. This is supported by strong blue shifts (upflow) measured in coronal and transition region emission lines \citep{harraMaterialOutflowsCoronal2001}. Coronal dimmings are therefore an important solar phenomenon, offering a direct window into the mass loss \citep{sterlingYohkohSXTObservations1997a} and magnetic reconfiguration linked to coronal mass ejections \citep{attrillUsingEvolutionCoronal2006}. An overview of dimmings can be found in \citet{veronigCoronalDimmingsWhat2025}. With a combination of reduced density and open-field structure, dimming regions resemble small, short-lived \citep[3--12\,hours; ][]{reinardCoronalMassEjectionAssociated2008} coronal holes and are therefore also known as transient coronal holes \citep{rustCoronalDisturbancesTheir1983}. 

One possible underlying cause of a dimming is the eruption of a quiescent filament \citep[e.g.][]{yangFormationTransientCoronal2009}. Quiescent filaments can be observed in H$\alpha$ and EUV emission as dark, thread-like structures in the quiet Sun. They consist of dense, cold plasma suspended high in the corona and are located above the polarity inversion line, which is the line between negative and positive magnetic fields in the photosphere \citep[e.g.][]{hirayamaModernObservationsSolar1985}. Quiescent filament eruptions are often accompanied by coronal mass ejections. \citet{jingRelationFilamentEruptions2004} reported that $54\%$ of quiescent filament eruptions are associated with coronal mass ejection and \citet{mccauleyProminenceFilamentEruptions2015} even reported $72\%$.

Despite the notable similarity between coronal holes and coronal dimmings, it is unclear to what extent the two phenomena have analogous plasma response. In this case study, we analyse a coronal dimming event, which formed after a filament eruption, with the aim of determining whether it temporarily transforms the quiet Sun into a coronal hole-like state.  A coronal hole-like state is, for example, characterised by significant plasma upflow in coronal emission lines, with formation temperatures of $\log(T/\text{K}) > 5.8$ \citep{tianNASCENTFASTSOLAR2010}. 

Spectroscopic observations and in situ solar wind measurements show that the elemental composition differs between coronal holes and the quiet Sun \citep[e.g.][]{feldmanCoronalCompositionSolar1998}. In particular, the abundances of elements with a low first ionisation potential (FIP $<10$ eV) are enhanced in the quiet-Sun corona, whereas coronal holes have elemental compositions that are closer to photospheric values. The general phenomenon in which low-FIP elements in the corona have enhanced abundances compared to the photosphere is called the FIP effect. High-FIP elements (FIP $\geq 10$ eV) do not show such enhancements. This effect is thought to be created by a fractionation mechanism in the chromosphere \citep[e.g. ][]{lamingUnifiedPictureFirst2004a}, in which ponderomotive forces generated by Alfvén waves preferentially act on ionized low-FIP elements, enhancing their transport into the corona relative to the predominantly neutral high-FIP elements. The fractionation process can progressively enrich low-FIP elements in closed magnetic loops, where the plasma is magnetically confined. The strength of the FIP effect is commonly measured with the FIP-bias, defined as the factor by which elements with a low first ionisation potential are enhanced in the corona relative to their photospheric abundances. Typical FIP-bias values reported for the quiet-Sun corona are of the order of $\sim 1.5$ \citep[e.g.][]{mihailescuWhatDeterminesActive2022, koCorrelationCoronalPlasma2016}. In coronal holes, plasma flows outward continuously; therefore, the elemental composition remains close to photospheric values \citep[i.e. FIP bias $\sim 1$; ][]{brooksEstablishingConnectionActive2011}.

In the chromosphere, it is difficult to distinguish coronal holes from the quiet Sun, but \citet{upendranFormationSolarWind2022} and \citet{kayshapDiagnosticsCoronalHole2015a} identified differences in Doppler velocities and intensities. When comparing areas with similar absolute magnetic flux density, $|B|$, chromospheric emission lines show enhanced up- and downflow and reduced intensities in coronal holes. 

In addition, EUV brightenings offer another observable relevant to understand the plasma properties in coronal holes and the quiet Sun. EUV brightenings are defined as a small, transient enhancement observed in EUV wavelengths. They comprise a range of phenomena, from the smallest events at the resolution limit of Solar Orbiter's High Resolution Imager (HRI) \citep[also called campfires; ][]{berghmansExtremeUVQuietSun2021} to bigger events as coronal bright points. Coronal bright points (CBPs) are small ($\lesssim 40$ Mm) loops or loop systems between flux concentrations of opposite magnetic polarities \citep[e.g.][]{madjarskaCoronalBrightPoints2019}. A recent overview of EUV brightenings in the quiet Sun and coronal holes can be found in \citet{harraDynamicsExtremeUltraviolet2025a}. \citet{subramanianCoronalHoleBoundaries2010} found using X-ray observations that brightenings are several times more frequent in coronal holes than in the quiet Sun. The occurrence is even higher along the coronal hole-quiet Sun boundary and around CBPs within the coronal hole. They also observed that coronal hole brightenings show more outflow, consistent with the idea that these brightenings are driven by reconnection between closed field lines and open field lines.

In this study, we investigate a dimming which followed a quiescent filament eruption. The objective is to determine whether the dimming event produced coronal hole-like behaviour in a quiet Sun region. The dimming was observed on 18 March 2022 during a coordinated Solar Orbiter-Hinode-IRIS campaign. The merging of this dimming with the southern coronal hole was studied by \citet{ngampoopunMergingCoronalDimming2023}. We investigate the similarity of the dimming to a coronal hole by analysing the coronal and chromospheric response prior to and during the dimming. This involves measuring the change in emission intensity, Doppler velocities, and FIP bias. We also investigate whether there are more EUV brightenings taking place at the dimming boundary; in principle, the region between the so-called open magnetic-field dimming region and closed magnetic-field quiet Sun. We also analyse whether CBPs that are initially located in a quiet-Sun region, which later undergoes dimming, show any response to the dimming event.

\section{Method}
We analyse observations on 17 and 18 March 2022  from the Solar Dynamics Observatory \citep[SDO;][]{pesnellSolarDynamicsObservatory2012}, Solar Orbiter \citep{mullerSolarOrbiterMission2020}, Hinode \citep{kosugiHinodeSolarBMission2007}, and from the Interface Region Imaging Spectrograph \citep[IRIS;][]{depontieuInterfaceRegionImaging2014}.

\subsection{Imaging data and line-of-sight magnetograms} 
From SDO, we used level-1 data of the 193 {\AA} and 171{\AA} channels of the Atmospheric Imaging Assembly \citep[SDO/AIA;][]{lemenAtmosphericImagingAssembly2012}. The images have a cadence of 12 seconds and an exposure time of 2 seconds. We also used line-of-sight (LOS) magnetograms from the Helioseismic Magnetic Imager \citep[SDO/HMI;][]{scherrerHelioseismicMagneticImager2012}. There are two types provided as part of the standard data products. The 45 s magnetograms are produced from filtergrams, while the 12 min magnetograms are computed from the Stokes vectors. We primarily used the latter given the lower noise \citep{liuComparisonLineofSightMagnetograms2012}. In some cases, the higher cadence was used, which we note explicitly. We calibrated the AIA data to level-1.5 either directly when downloading it with JSOC \footnote{\url{http://jsoc.stanford.edu/}} or afterwards with aiapy \citep{Barnes2020}.

We used two 1-hour level-2 datasets from Solar Orbiter's High Resolution Imager at 174 {\AA} \citep[HRI\textsubscript{EUV};][]{rochusSolarOrbiterEUI2020}. The observations were taken on 17 March 2022 09:47-10:47~UT and 18 March 2022 10:10-11:10~UT, when Solar Orbiter was at heliocentric distances of 0.38 AU and 0.37 AU, with angular separations from  Earth of $28^\circ$ and $32^\circ$, respectively. The cadence of the datasets is 5 s and the exposure time is $2.8 \text{ s}$. The jitter in the HRI\textsubscript{EUV} datasets was removed by maximizing the cross-correlation according to \citet{chittaSolarCoronalHeating2022}. 

To maintain a co-alignment between the HRI\textsubscript{EUV} and the SDO observations, we used additional level-2 data from  Solar Orbiter's Full Sun Imager \citep[FSI;][]{rochusSolarOrbiterEUI2020} in the 174 {\AA} and 304 {\AA} passband and level-1 SDO/AIA 304 {\AA} data. First, we determined the spatial offset between HRI\textsubscript{EUV} and FSI\textsubscript{174}. Second, we computed the offset between FSI\textsubscript{304} and AIA 304~{\AA}. Both offsets were calculated using the SunPy \texttt{sunkit\_image.coalignment} module \citep{sunpy_community2020}. By combining these two offsets, we obtained the total shift required to align HRI\textsubscript{EUV} with all AIA channels. This approach relies on the assumption that level-2 FSI 304 {\AA} and 174 {\AA} are already well aligned and that level-1.5 AIA 304 {\AA} is also well aligned to the other AIA channels. The alignment precision achievable with this method is expected to reach AIA-pixel level (i.e. 0.6\arcsec). 

\subsection{Spectroscopic data}
\begin{figure*}
    \centering
    \includegraphics[width=0.8\linewidth]{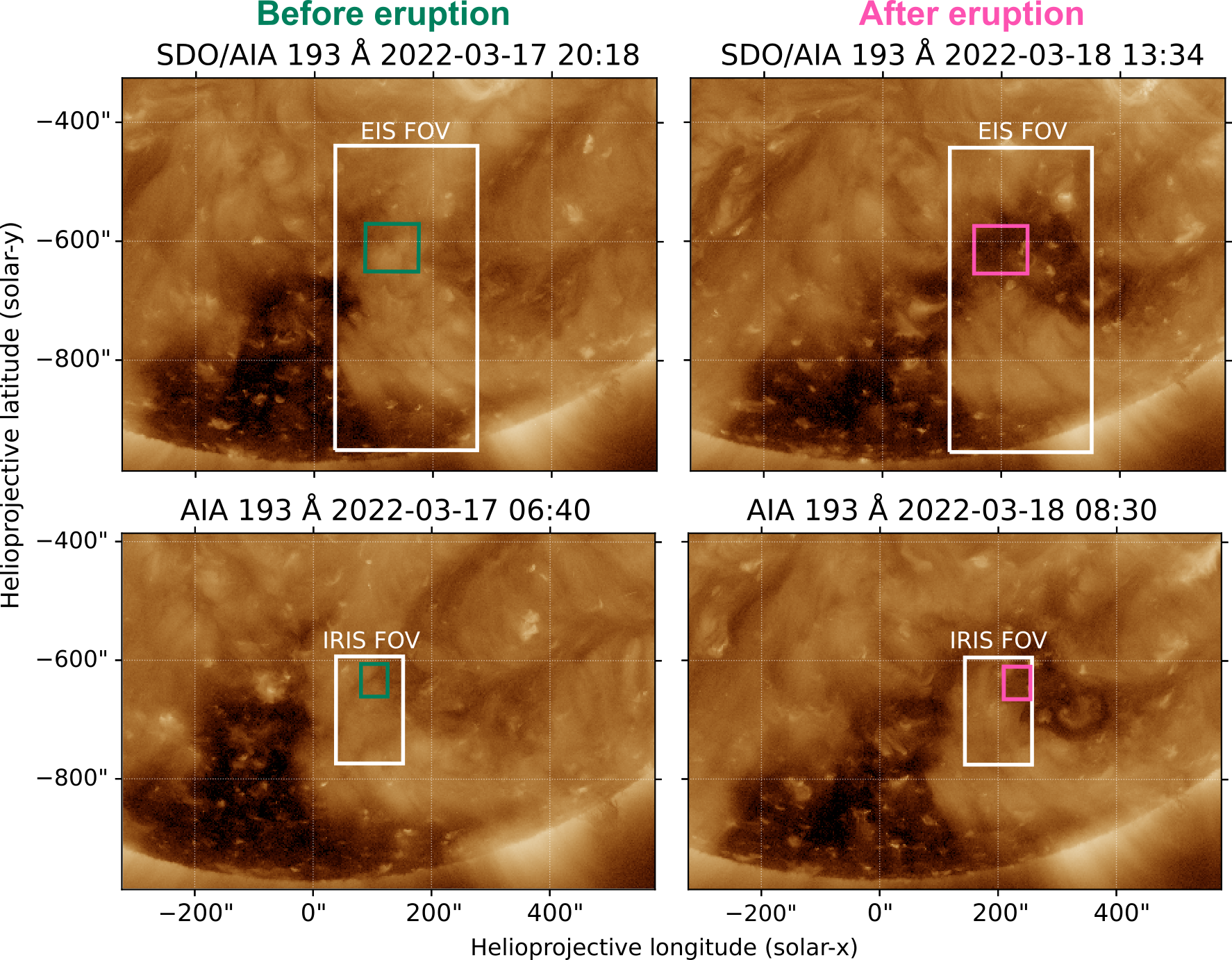}
    \caption{Raster fields of view (FOVs) before and after the eruption. The white rectangles show the FOVs of the two IRIS and two Hinode/EIS rasters, plotted on the SDO/AIA 193~{\AA} images at the central time of each raster. The green and pink rectangles indicate the areas used to compare the conditions before and after the eruption, they are selected such that they lie entirely within the dimming region after the eruption.}
    \label{fig:FOVs_raster}
\end{figure*}

The dimming event was also observed with Hinode's \citep{kosugiHinodeSolarBMission2007} EUV Imaging Spectrometer \citep[EIS;][]{culhaneEUVImagingSpectrometer2007a} and IRIS \citep{depontieuInterfaceRegionImaging2014}. These two spectrometers were built to investigate energy transport throughout the solar atmosphere. Hinode/EIS covers wavelengths corresponding to the upper transition region and corona, while IRIS focuses on the chromosphere and transition region. Together, they allow for better understanding across the various layers of the atmosphere. From both instruments, we used two spectral rasters, one taken before the filament eruption and one after the dimming  formed. Their field of view (FOV) is plotted with a white rectangle on top of SDO/AIA 193~{\AA} images in Fig. \ref{fig:FOVs_raster}.

 We used two 120-step raster scans covering a field-of-view of 240\arcsec$\times$512\arcsec from Hinode/EIS. The raster times are 17 March 2022 18:35 to 21:59~UT and on 18 March 2022 11:52 to 15:16~UT. At each slit position, the exposure time was 100 s.

From the IRIS spectrograph, we used two level-2 very large dense 320-step rasters, recorded on 17 March 2022 from 06:15 to 07:06~UT and on 18 March 2022 from 08:05 to 08:55~UT. At each slit position there was an exposure time of 8 seconds and a step cadence of 9.5~s or 9.3~s. The latter raster used 2$\times$ spatial binning along the slit and 2$\times$ spectral binning. The main spectral lines used in this study and their corresponding ions and temperatures are listed in Table \ref{table:spectral_lines}.
\begin{table}[ht!]
    \caption{List of spectral lines used in this work.}                
    \label{table:spectral_lines}  
    \centering                       
    \begin{tabular}{l c c}      
        \hline\hline             
        Ion & Wavelength  & Height / Temperature log(T[K]) \\         
         & [{\AA}] & \\
        \hline                    
            \ion{Fe}{xii} & 195.1 & corona / 6.2 $^*$\\  
            \ion{Si}{x} & 258.4 & corona / 6.1 $^*$\\ 
            \ion{S}{X} & 264.2 & corona / 6.2 $^*$\\
            \ion{Mg}{ii} k\textsubscript{3} & 2796 & upper chromosphere $^{**}$  \\
            \hline 
    \end{tabular}
    \tablefoot{ $^{*}$ Peak temperatures of the contribution function of respective atomic species. We assumed an electron density of $10^9$ cm$^{-3}$ and extracted the contribution function from the CHIANTI database \citep{dere2023chianti} Version 11.0.2 with ChiantiPy. $^{**}$ Formation region according to \citet{leenaartsFormationIRISDiagnostics2013a}.}
\end{table}

\subsubsection{Preprocessing Hinode/EIS data}\label{sec:eis}
The EIS data were processed using the standard calibration software (eis\_prep) available in SSW \citep[SolarSoftware,][]{Freeland1998}. The routine subtracts the CCD dark current, cleans warm, hot, dusty pixels, and cosmic ray hits, and converts the data to physical units. It also corrects for the drift of the spectrum across the CCD, due to the thermal environment of the instrument around the satellite orbit, using the artificial neural network (ANN) model implemented by \citet{Kamio2010}. Spectra taken on the short- and long-wavelength detectors are also spatially aligned. 

There is no calibration lamp on board EIS or photospheric emission lines to provide an absolute wavelength scale, so we employed the relative Doppler velocities as the measured velocities. We assigned the reference wavelength by averaging a quiet Sun area. For these observations, we used the top 52 pixels of the raster. This area is clearly a quiet Sun region in the first raster and is still primarily dominated by quiet Sun post-eruption. 

The ANN model was developed early in the mission and is less accurate when applied to recent data. We identified a residual orbital spectral drift across the raster in these datasets and removed it using a 5-pixel smoothed average of the quiet Sun region along the slit. The ANN root mean square error is $\sim$4.5 km/s and the assumption of zero velocity at the formation temperature of \ion{Fe}{xii} 195.119~\AA\ might underestimate the blueshift by a few km/s \citep{Peter1999}. We fit a double Gaussian function to the \ion{Fe}{xii} 195~\AA\ spectral profile to account for the density sensitive blend at 195.179~\AA. 

We used the \ion{Si}{X} 258.4~{\AA} / \ion{S}{X} 264.2~{\AA} intensity ratio as a proxy for the FIP bias. These lines are formed at similar temperatures and densities and are close in wavelength. We refer to this ratio simply as the FIP bias. The ratio has been widely used in EIS elemental abundance studies, including a full analysis that accounts for temperature and density effects, and a comprehensive examination of pixel-scale maps of the entire solar disk \citep{Brooks2015}. Broadly speaking, this method provides a FIP proxy that non-linearly compresses the range of the underlying FIP bias. 

The EIS photometric calibration uncertainty was estimated in the laboratory on the ground to be $\sim$23\% \citep{Lang2006}. This yields an uncertainty of $\sim$30\% on the intensity ratio used for the FIP bias proxy. We fit single Gaussian functions to the \ion{Si}{X} 258.4~{\AA} and \ion{S}{X} 264.2~{\AA} profiles.

\subsubsection{Preprocessing the IRIS data}\label{sec:iris}
We retrieved IRIS level-2 data files from the Heliophysics Event Knowledgebase \citep[HEK;][]{hurlburtHEK2012}. We performed the radiometric calibration and despiked the level-2 data. We then extracted the intensities and Doppler shifts of the \ion{Mg}{ii} k\textsubscript{3} feature using the \texttt{iris\_get\_mg\_features} procedure \citep{pereiraExtractMgIIFeatures2013} available in SSW. The 17 March IRIS dataset was rebinned by a factor of two along both the slit and spectral axes to match the onboard binning of the 18 March dataset. Doppler velocity values were corrected for thermal and orbital drift by fitting the photospheric lines.

We performed a Monte Carlo simulation with 100 realizations. For each realization, we re-extracted Mg\,\textsc{ii} features from the IRIS spectra with randomly generated noise. The noise at each detector pixel was drawn from a Gaussian distribution with a standard deviation given by the combined photon shot noise and CCD readout noise. We used the median of the k\textsubscript{3} intensities and Doppler shifts across the 100 Monte Carlo realizations, taking the corresponding standard deviations as the measurement uncertainties. The Doppler velocities have on average a measurement error of $1.3\,\mathrm{km\,s^{-1}}$ (first raster, 17 March 2022 06:40 UT) and $1.5\,\mathrm{km\,s^{-1}}$ (second raster, 18 March 2022 08:30 UT) and intensities of $4.4\times 10^3$ and $4.1 \times 10^3  \ \mathrm{erg\,s^{-1}\,cm^{-2}\,sr^{-1}\,\text{\AA}^{-1}}$, respectively.  

\subsection{Generating pseudo-rasters}\label{sec:pseudo-raster}
We generated HMI pseudo-rasters that correspond to the IRIS raster. To this end, we followed the method described in \citet{upendran2021properties}. From IRIS, we used the slit-jaw images (SJI) at $2796\;\text{\AA}$ and the AIA $1600\;\text{\AA}$ images from SDO, along with HMI LOS magnetograms with a 45\;s cadence. First, we aligned the SJI and IRIS raster by identifying the slit position in SJI (visible as dark line) and then shifting it to the corresponding raster position. This is justified for the IRIS observations used because the SJI cadence is exactly half of the raster-step cadence. Consequently, each SJI image has a corresponding raster position that is observed simultaneously. Then we downsampled the SJI and HMI images to the AIA resolution and co-aligned the AIA and SJI images on a sub-pixel scale by maximizing the cross-correlation. The same shift was then also applied to the HMI images. The next step was to reproject the AIA and HMI images to the IRIS raster FOV. To create the pseudo-rasters, we extracted, at each raster step, the corresponding AIA and HMI data from the map whose observation time is closest to that step.

\subsection{Coronal dimming contour}
Following \citet{dissauerDetectionCoronalDimmings2018}, we defined a reduction in intensity of 35\% in SDO/AIA 193~{\AA} images as the dimming region. To follow this dimming region in time, we rotated each image to the same reference frame using SunPy's context manager \texttt{sunpy.coordinates.propagate\_with\_solar\_surface} \citep{sunpy_community2020}. We used the eruption time (18 March 2022, 02:00 UT) as the reference frame. For each frame, we computed a base-ratio image by dividing the rotated frame by the reference image and defined the dimming as the longest connected contour (measured by a circumference) obtained from a 0.65 threshold. If the second-largest contour was at least half of the largest contour and within a distance of 5\arcsec, we combined the two contours for the dimming area. This helps avoid artificial disappearance and reappearance of parts of the dimming region between frames when the region briefly splits into two pieces.

\subsection{Detection of EUV brightenings}\label{sec:detection}
We detected EUV brightenings automatically using a wavelet-based algorithm \citep{berghmansExtremeUVQuietSun2021}. This algorithm transforms each frame into a set of wavelet scales, where each scale isolates intensity variations on a specific spatial size (from fine to coarse). In each frame, we estimated the noise as $\sigma = \sqrt{\text{(read out noise)}^2 + \text{(shot noise)}^2}$ and marked the pixels as significant if the signal in one of the first $N_\text{levels}$ wavelet bands exceeds $k_\sigma\cdot \sigma$. The threshold, $k_\sigma$, effectively sets the detection sensitivity of the algorithm. To select this value, we performed the detection several times with different $k_\sigma$ values and visually inspected the results. We required obvious brightenings to be detected, while keeping false detections (especially at the dimming border) as low as possible. The number of wavelet bands, $N_\text{levels}$, sets the largest spatial scale considered in the detection. Increasing $N_\text{levels}$ also allows for the detection of larger brightenings. Additionally, we used the requirement that all brightenings must span at least two pixels in at least one frame and persist for at least two consecutive frames. This eliminates the detection of cosmic rays or faulty instrument pixels.

With HRI\textsubscript{EUV} images, we used a threshold of $k_\sigma=8.6$ and $N_\text{levels}=2$. This allows us to detect the smallest EUV brightenings, also known as campfires \citep{berghmansExtremeUVQuietSun2021}. For detections in the SDO/AIA data, we used one additional wavelet scale, i.e., $N_\text{levels}=3$. Furthermore, we used a threshold of $k_\sigma=5\sigma$ for SDO/AIA 193~{\AA} and $k_\sigma=6\sigma$ for SDO/AIA 171 {\AA}.

We omitted the step of transforming the images into Carrington coordinates before applying the detection, since the region of interest is close to the limb, where remapping would introduce strong distortion. However, this meant that the detected brightenings are always a projection onto the image plane.

The strength of the wavelet-based detection algorithm and its application have been demonstrated in other studies \citep[e.g.][]{berghmansExtremeUVQuietSun2021, barczynskiStatisticalComparisonEUV2022, narangExtremeultravioletTransientBrightenings2025, limQuasiperiodicPulsationsExtremeultraviolet2025}. However, there are still some limitations that should be considered when interpreting the results. We adapted the code and used it for the first time in a dimming region, which could lead to some further limitations. 

The detection algorithm has some unresolved limitations when used for the statistical analysis of EUV brightening occurrence rates. It considers image pixels significant only if they exceed $k_\sigma$ times the standard deviation of the noise. This means that brightenings that are too faint and do not pass this strict threshold will not be included. Furthermore, it could be that a brightening falls below the threshold for a few frames, this leads to an interruption in detection and is then interpreted as two separate brightenings. This means that with a low threshold, we can detect one continuous brightening; however, with a high threshold, it can be detected as multiple brightenings, one after the next. Using an additional wavelet scale, we also increased the probability that two nearby brightenings are detected together as one brightening. 

In SDO/AIA 193~\AA\, we found with a low threshold ($k_\sigma=3\sigma$) that the sharp intensity step between the quiet Sun and dimming does cause detections along the dimming border, which we would not manually classify as EUV brightenings. We were able to remove these faulty detections by increasing the threshold to $k_\sigma=5\sigma$. 

\subsection{Magnetic field modelling}\label{subsec:MagneticModelling}
To obtain the  coronal magnetic configuration (i.e., field line distribution, field line heights and lengths) at the location of the CBPs, we extrapolated the 12 min LOS photospheric magnetic field as observed by HMI. The magnetic field was computed in the potential approximation: $\vec{\nabla} \times \vec{B}= 0$ \citep[see][and references therein for a description of the model and its limitations]{Mandrini15}. The model was first transformed from the local frame to the observer frame, as discussed in \citet{Mandrini15}. 

\section{Results}
\subsection{Data overview}

\begin{figure}
    \centering
    \includegraphics[width=\linewidth]{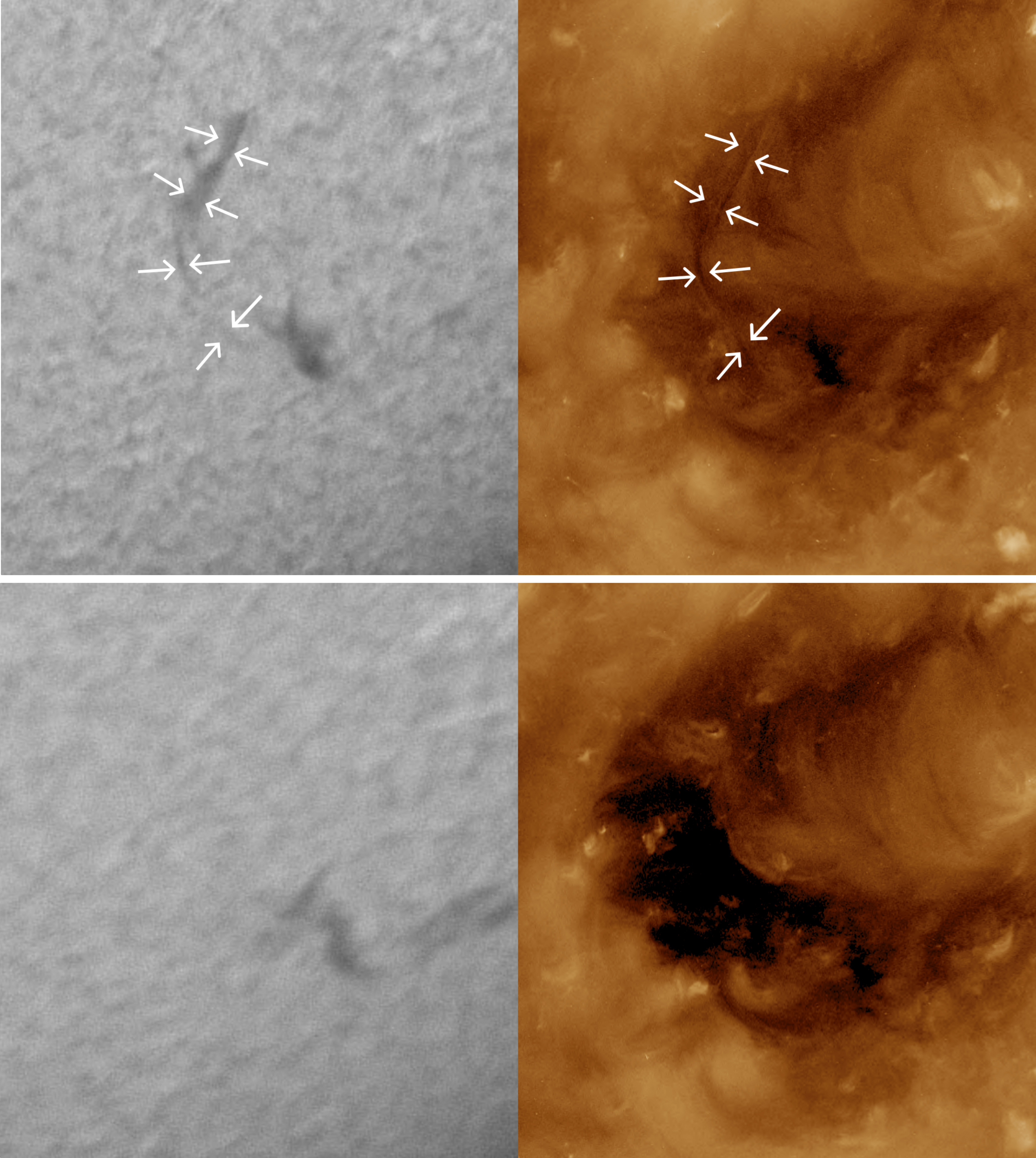}
    \caption{Filament eruption in the dimming region. The pre- and post-filament-eruption states are depicted in the top and bottom rows, respectively, at 01:19 and 04:19 UT on 18 March 2022. The left column shows H$\alpha$ observations from the Global Oscillation Network Group (GONG, National Solar Observatory). The right column SDO/AIA 193~{\AA} images of the same FOV. The filament appears as a dark, band-like absorption feature between the white arrows. The coronal dimming can be seen in the bottom panel, as a new dark region in SDO/AIA 193 \AA. The images were generated with JHelioviewer \citep{mullerJHelioviewerTimedependent3D2017}. The associated movie is available online.}
    \label{fig:filament_eruption}
\end{figure}

\begin{figure}
    \centering
    \includegraphics[width=1\linewidth]{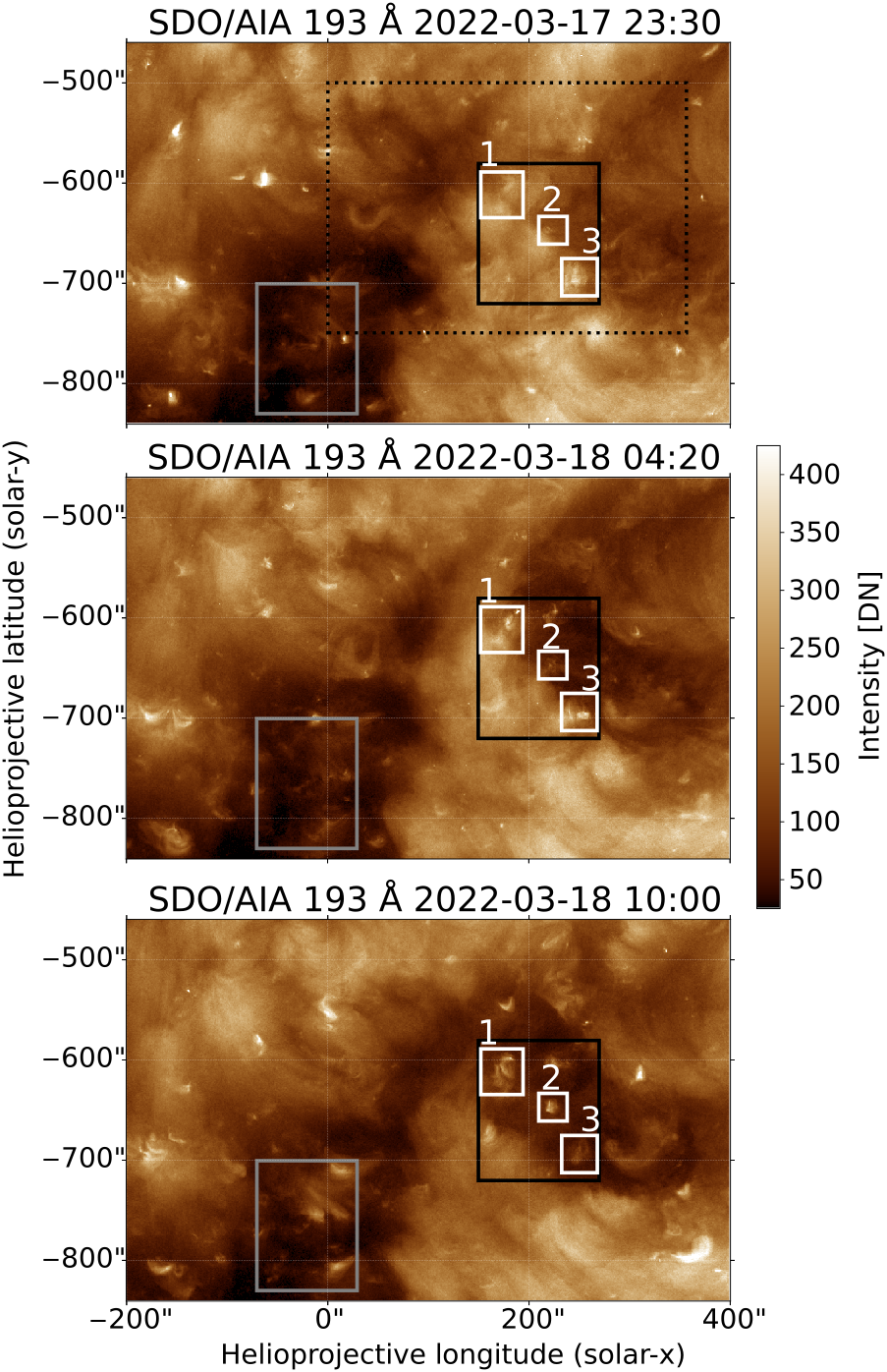}
    \caption{Evolution of the coronal dimming and the southern coronal hole with marked analysis regions. The SDO/AIA 193~\AA\ images are ordered in time from top to bottom, the filament eruption occurred between the first and second image. The gray-dashed rectangle shows a region within the southern coronal hole and the black-solid rectangle shows a region where the dimming forms. The white rectangles show the area of three selected coronal bright points. The black-dotted rectangle in the upper image corresponds to the same FOV as plotted in Fig.~\ref{fig:BPs_mag_modelling} for the magnetic models of the coronal bright points.}
    \label{fig:areas}
\end{figure}

\begin{figure}
    \centering
    \includegraphics[width=1\linewidth]{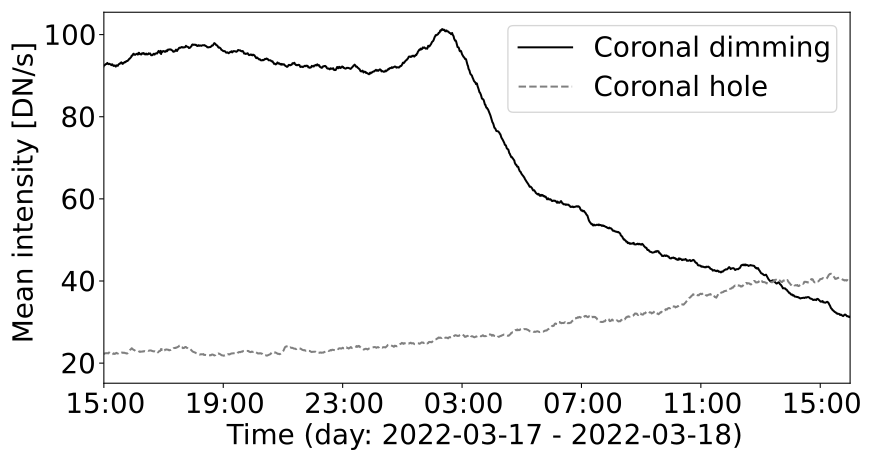}
    \caption{SDO/AIA 193 {\AA} mean intensity evolution. The mean intensity is measured between 17 March 2022 15:00 UT and 18 March 2022 16:00 UT inside a coronal hole (gray-dashed) and within a region where the dimming forms (black-solid).}
    \label{fig:intensity_time}
\end{figure}

A quiescent filament erupted at $\sim$ 02:00 UT on 18 March 2022 and led to the formation of a coronal dimming region. The filament eruption and onset of coronal dimming are shown in Fig.~\ref{fig:filament_eruption}. In relation to this event, a coronal mass ejection was observed. According to the SOHO/LASCO CME catalogue \footnote{\url{https://cdaw.gsfc.nasa.gov/CME_list/}}, the coronal mass ejection had its first LASCO/C2 appearance at 03:24 UT with a linear speed of 256 km/s. 
The temporal evolution of the coronal dimming is illustrated at three time steps in SDO/AIA~193~{\AA} observations in Fig.~\ref{fig:areas}. The rectangles define areas used for the analysis. The mean intensities within the dimming and coronal hole regions (black and gray rectangles, respectively) are shown in Fig.~\ref{fig:intensity_time}. The temporal evolution of the mean intensity within the dimming region exhibits a maximum at 02:20 UT followed by a decrease. This behaviour corresponds to the passage of the EUV wave, which is observed as a propagating bright front (intensity peak), followed by an expanding dimming region (subsequent intensity reduction). The mean intensity in the coronal hole increases over time, which is due to new structures emerging and LOS effects. The dimming area reaches within 11 hours a mean intensity of 40 DN/s, which matches, at this time, the average intensity in the coronal hole. This corresponds to a reduction of 55-60$\%$ from the intensity before the EUV wave. 

\begin{figure}
    \centering
    \includegraphics[width=\linewidth]{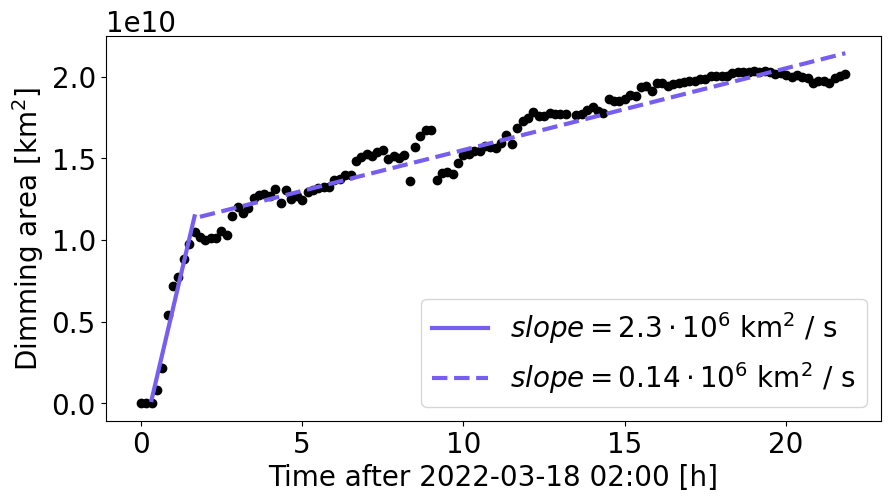}
     \caption{Area of the dimming region plotted against the time after the eruption. The data is fitted linearly between 20 and 100 minutes after 02:00 UT and also between 1.8 to 21.8 hours. The online movie shows the evolution of the dimming contour, which we used to infer the dimming area. The associated movie is available online. }
    \label{fig:grow_rate}
\end{figure}

The growth of the dimming area can be divided into two parts: the initial phase with fast growth and the subsequent phase with slow growth (see Fig.~\ref{fig:grow_rate}). The initial phase lasts about 80 minutes and has a growth rate of $(2.3\pm 0.2) \cdot 10^6 \text{ km}^2\text{s}^{-1}$, while the growth rate for the next 20 hours is  $(0.14 \pm 0.01) \cdot 10^6 \text{ km}^2\text{s}^{-1}$. The errors represent the $1\sigma$ uncertainty of the fitted coefficient.

\subsection{Spectroscopic response of the dimming}
We compared the spectroscopic plasma properties of the IRIS and Hinode/EIS rasters in the same area before the eruption (quiet Sun) and after the eruption (dimming). This corresponds to the green and pink boxes in Fig.~\ref{fig:FOVs_raster}.

\begin{figure}
    \centering
    \includegraphics[width=\linewidth]{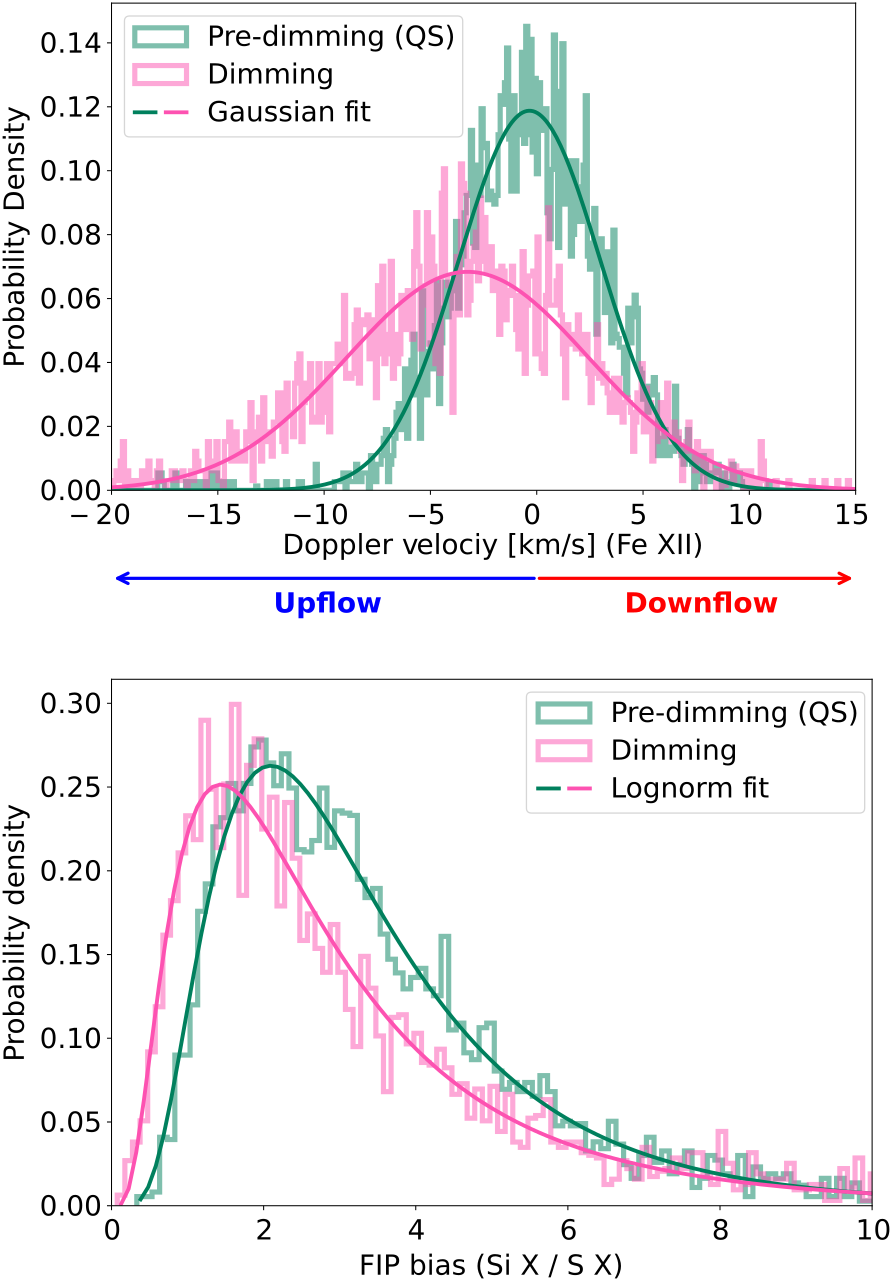}
    \caption{Doppler velocity (top) and FIP-bias (bottom) measured with Hinode/EIS. Values derived from the pre-dimming raster (17 March 2022, 20:17 UT) are plotted in green and values from the raster taken after the eruption inside the dimming (18 March 2022, 13:34 UT) in pink. }
    \label{fig:EIS_histograms}
\end{figure}

We analysed the Hinode/EIS coronal emission line \ion{Fe}{xii} 195.119~{\AA} to quantify the LOS plasma flows. First, we prepared the data and fitted a double Gaussian in every pixel as described in Sect.~\ref{sec:eis}. This procedure yields the Doppler velocity for each pixel located within the pink and green rectangles, respectively (Fig. \ref{fig:FOVs_raster}). The scan across the width of this selected region required 77 minutes. The resulting distributions of pixel-wise Doppler velocities are shown in the histogram in Fig.~\ref{fig:EIS_histograms} (top). We can see that the distribution gets broader and there is more upflow overall after the eruption. The peak of the histogram is determined from a single Gaussian fit and changes from $-0.34^{+0.37}_{-0.27}\,\mathrm{km\,s^{-1}}$ to $-3.2^{+0.4}_{-0.6}\,\mathrm{km\,s^{-1}}$. 

We calculated for each pixel the FIP-bias proxy (\ion{Si}{X} 258.4~{\AA} / \ion{S}{X} 264.2~{\AA} intensity ratio), following the procedure outlined in Sect.~\ref{sec:eis}. The resulting distribution is shown in the lower panel of Fig.~\ref{fig:EIS_histograms}. We observe that the FIP bias in the dimming region is lower compared to the value measured in the quiet Sun prior to the eruption. The peak of a log-normal fit decreases from $2.1^{+0.1}_{-0.1}$ to $1.4^{+0.1}_{-0.1}$.

The uncertainties denote the $95\%$ sampling confidence intervals estimated using a 10$^4$-step spatial block bootstrap. The Doppler and intensity maps are resampled by drawing contiguous spatial blocks with replacement. The blocks have dimensions of $l_x \times l_y$, where $l_x$ and $l_y$ correspond to the spatial correlation lengths in the $x$ and $y$ directions. These were estimated as the distance at which the 2D autocorrelation function drops below $1/e$.

\begin{figure}
    \centering
    \includegraphics[width=0.7\linewidth]{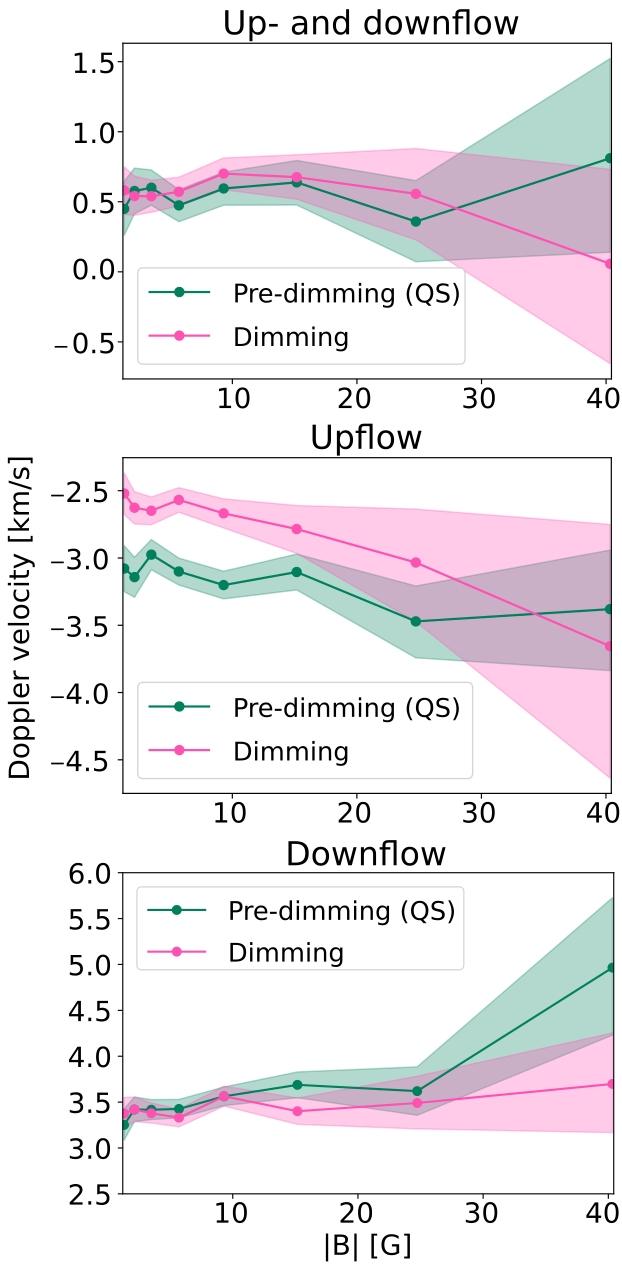}
    \caption{\ion{Mg}{ii} k$_3$ mean Doppler velocities measured with IRIS and binned with the absolute LOS magnetic field strength  measured with SDO/HMI. 
    The shaded regions represent the 95\% uncertainty interval, estimated with bootstrap resampling. The values are derived from the same subsection, covering the quiet Sun in the raster from 17 March 2022 at 06:40 UT (green) and the coronal dimming in the raster from 18 March 2022 at 08:30 UT (pink). }
    \label{fig:DopplerBinnedWithB}
\end{figure}

We investigated upper-chromospheric plasma flows using Doppler velocities from the IRIS \ion{Mg}{ii} k$_3$ spectral line. The Doppler velocity in each pixel is measured following the procedure outlined in Sect.~\ref{sec:iris}. We also construct pseudo-rasters from SDO/HMI LOS magnetograms that are co-spatial and co-temporal with the IRIS rasters (see Sect~\ref{sec:pseudo-raster}). This approach provides, for every raster pixel, a corresponding Doppler velocity and magnetic field strength. We only considered pixels within the dimming subsection of the raster, marked by the pink and green rectangles in Fig. \ref{fig:FOVs_raster}). For both rasters, these pixels were recorded within 20 minutes. Similarly to the approach taken by \citet{upendranFormationSolarWind2022}, we sorted all the pixels into bins based on their magnetic field strength and computed, for each bin, the mean Doppler velocity over all pixels, all pixels with upflow ($v < 0\ \mathrm{km\ s^{-1}}$) and all pixels with downflow ($v > 0\ \mathrm{km\ s^{-1}}$). The resulting relationships between Doppler velocity and magnetic field strength are shown in Fig.~\ref{fig:DopplerBinnedWithB}. In the quiet Sun up to $\sim 0.5\,\mathrm{km\,s^{-1}}$, we found a stronger upflow compared to the dimming; otherwise, there are no significant changes (see Fig.~\ref{fig:DopplerBinnedWithB}). 

\begin{figure}
    \centering
    \includegraphics[width=\linewidth]{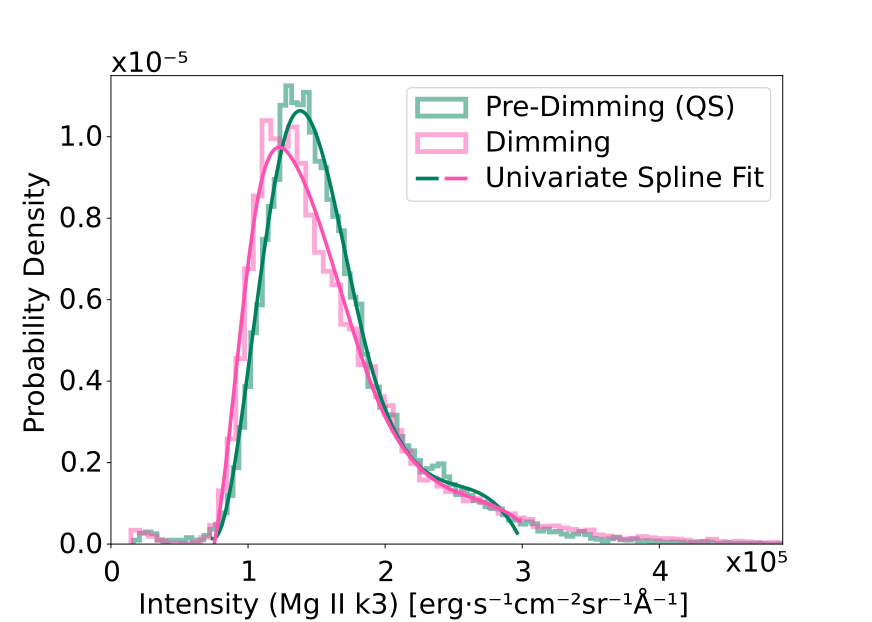}
    \caption{Spectral radiance measured with the \ion{Mg}{ii} k3 emission line from IRIS. The values are derived from the same subsection, covering the quiet Sun in the raster from 17 March 2022 at 06:40 UT (green) and the coronal dimming in the raster from 18 March 2022 at 08:30 UT (pink). }
    \label{fig:IRIS_histogram}
\end{figure}

The intensity in each pixel within the \ion{Mg}{ii} emission line, as observed with IRIS, was calculated following the methodology described in Sect.~\ref{sec:iris}. The intensity distribution is plotted in Fig.~\ref{fig:IRIS_histogram}. We can see lower intensities in the dimming compared to the quiet Sun. We used a Univariate Spline to identify the peak position, which was reduced by $11\%$, with a $95\%$-confidence interval of $[-16.7,\,-0.6]\%$. The intensity decreases from $1.38^{+0.03}_{-0.11}\times 10^5 $ to $1.23^{+0.08}_{-0.09}\times 10^5 \ \mathrm{erg\,s^{-1}\,cm^{-2}\,sr^{-1}\,\text{\AA}^{-1}}$.

\subsection{EUV brightenings}
\begin{figure}
    \centering
    \includegraphics[width=\linewidth]{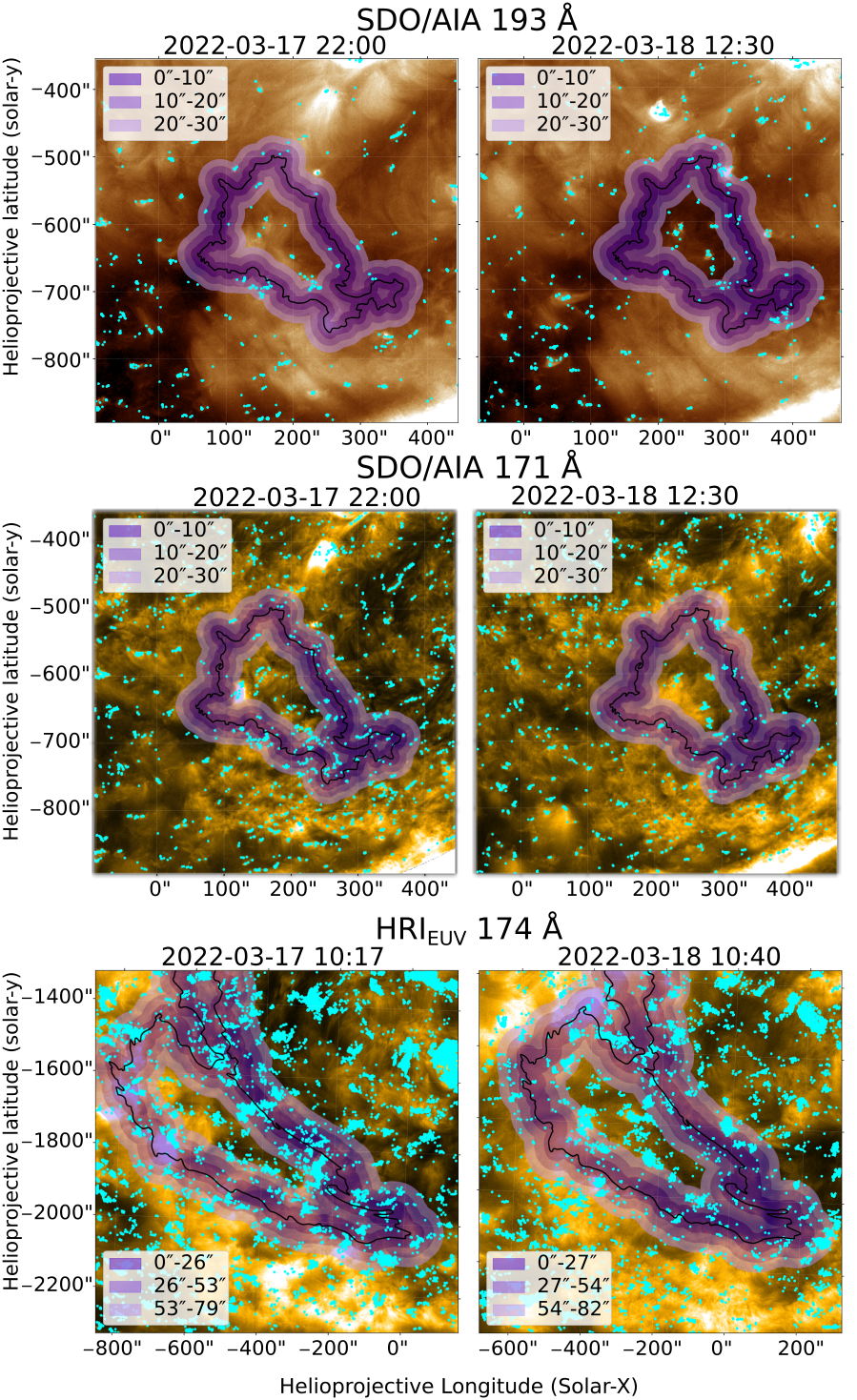}
    \caption{Detected EUV brightenings in SDO/AIA 193 {\AA} (top), SDO/AIA 171 {\AA} (middle) and HRI\textsubscript{EUV} 174 \AA\ (bottom). The images are at mid-time of the corresponding detection range. The brightenings were detected over 2 hours for SDO/AIA and over 1 hour for HRI\textsubscript{EUV}. Each cyan dot represents one EUV brightening and is plotted at the centre position. The black line shows the contour of the dimming, which was detected on the corresponding SDO/AIA 193 {\AA} image at 12:30 or 10:40 on 18 March 2022. The purple areas around the dimming contour correspond to distances of $0''-10''$, $10''-20''$, and $20''-30''$ when viewed from Earth. 
    }
    \label{fig:brightenings_in_distance}
\end{figure}

In this section, we aim to measure whether the number of EUV brightenings increases or remains unchanged when the dimming forms, with an emphasis on the vicinity of the dimming boundary. We detected (see Sect.\ref{sec:detection}) all brightenings in SDO/AIA 193 {\AA} and SDO/AIA 171 {\AA} during five different time intervals: one before the eruption (17 March 2022 21:00-23:00) and four after the eruption, when the dimming was present (18 March 2022 07:30-09:30, 09:30-11:30, 11:30-13:30 and 13:30-15:30). For each post-eruption interval, we used the dimming contour defined at the middle of that time-interval. We then derived the brightening density as a function of distance from the contour and compared the post-eruption density to the pre-eruption density in the same area. The distance to the contour is defined as the minimum Euclidean distance between the brightening (centre) and any point on the contour. An example of detected brightenings can be seen in Fig.~\ref{fig:brightenings_in_distance}.

\begin{figure}
    \centering
    \includegraphics[width=0.8\linewidth]{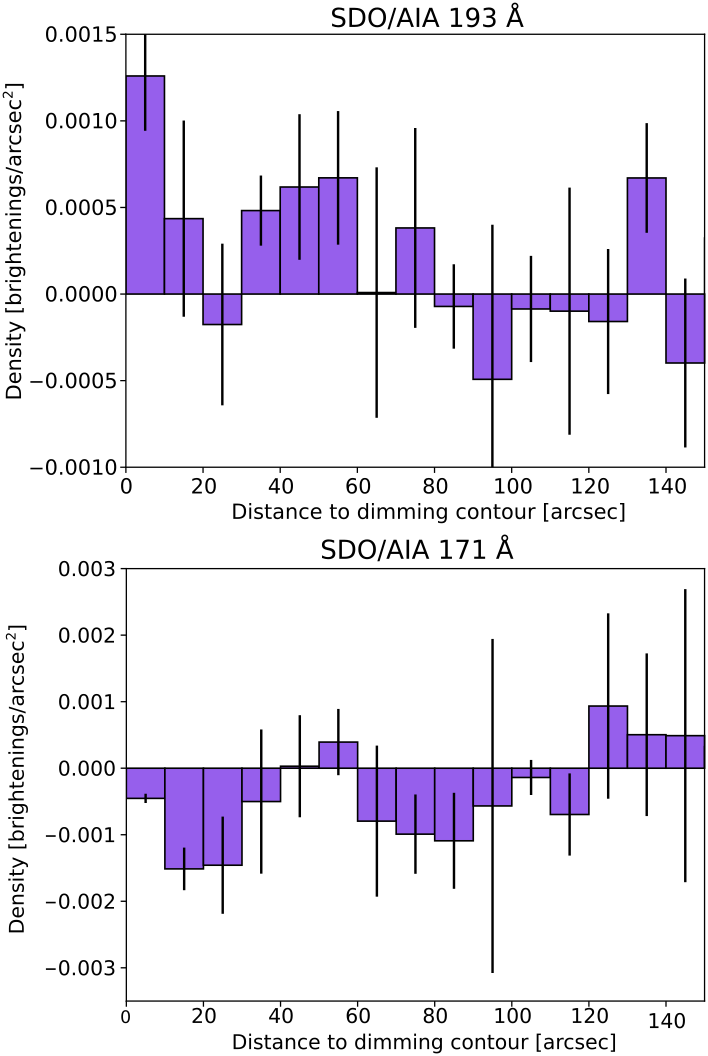}
    \caption{Difference in EUV-brightenings between pre- and post-eruption intervals relative to the dimming contour. The results are shown for SDO/AIA 193~{\AA} (top) and SDO/AIA~171 {\AA} (bottom) in 10\arcsec bins. The mean density difference is plotted for the four post-eruption time-intervals, and the black bars indicate the standard deviation.}
    \label{fig:histogram_brightening193}
\end{figure}

In SDO/AIA 193 {\AA}, after the eruption, we find an average of $(29\pm6)\%$ more brightenings compared to the pre-eruption interval. The average was taken over the four post-eruption intervals and the error represents the standard deviation. Inside the dimming, a similar increase by $(40\pm30)\%$ is seen. However, we find a more pronounced enhancement in the number of EUV brightenings close to the dimming border (see Fig.~\ref{fig:histogram_brightening193}). Within $10\arcsec$ around the dimming border (dark purple area in Fig.~\ref{fig:brightenings_in_distance}), we find, on average, $3.5 \pm 0.3$ times more brightenings compared to the same area in the reference interval before the eruption. $(75 \pm 9) \%$ of these brightenings close to the contour are outside the dimming. 

\begin{figure}
    \centering
    \includegraphics[width=\linewidth]{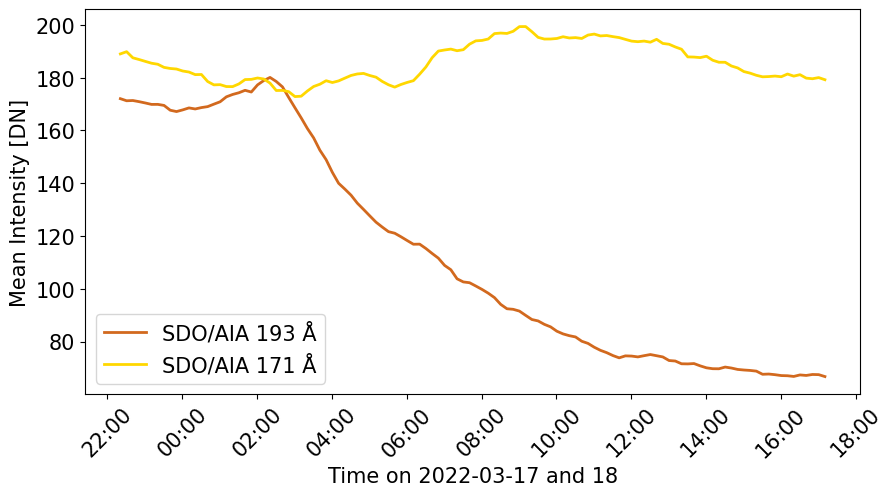}
    \caption{Temporal evolution of the SDO/AIA {193~\AA} and {171~\AA} channel intensities. The intensities are averaged over the same region between 22:20 UT on 17.03.2022 and 17:20 UT on 18.03.2022.}
    \label{fig:intensity}
\end{figure}

Repetition of the same measurement, but for SDO/AIA 171~{\AA} instead of SDO/AIA 193 {\AA} did not result in a brightening enhancement close to the dimming border (see Fig.~\ref{fig:histogram_brightening193}, bottom). In general, there is a decrease in the number of brightenings by $(3\pm 4)\% $ , which is mainly due to the fact that there are $(40\pm10)\% $ fewer brightenings within the dimming contour. We think this reduction might be due to an enhanced intensity in the 171~\AA channel (see Fig.~\ref{fig:intensity}), which makes it more difficult to detect brightening above the bright background. The intensity is measured within the dimming contour identified at 11:30 UT on 18.03.2022 propagated to each time step by accounting for solar rotation. 

\begin{figure}[!ht]
    \centering
    \includegraphics[width=0.8\linewidth, trim=0 0 0 0cm, clip=true]{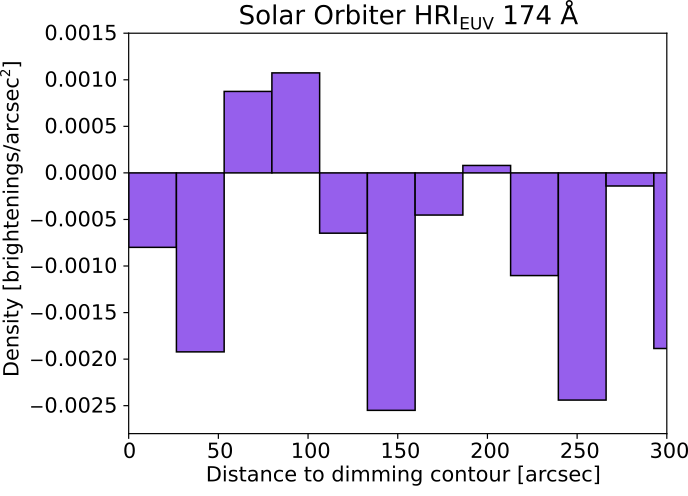}
    \caption{Same as Fig.~\ref{fig:histogram_brightening193}, but with EUV brightenings detected in HRI\textsubscript{EUV} 174 {\AA}. The bin size is 27\arcsec of the Solar Orbiter view, which corresponds to 10\arcsec viewed from Earth.}
    \label{fig:histogram_brightening174}
\end{figure}

In Solar Orbiter's high resolution 174~\AA\ images, we detected smaller brightenings, also called campfires. Due to the limited temporal coverage of the HRI\textsubscript{EUV} only one pre-eruption and on post-eruption interval (each one hour) was available. Therefore, no standard deviation is calculated for these measurements. We found that the number of brightenings decreases in the whole FOV after the eruption by 28\% and inside the dimming by $9\%$. Similarly to SDO/AIA 171 {\AA}, the brightening density is not enhanced close to the dimming border (Fig.~\ref{fig:histogram_brightening174}).

\subsection{Coronal bright points}
In this section, we describe how we analysed the CBPs, located inside the region that dims. We compare them before, during and after the eruption to determine whether their activity changes. We selected three CBPs (Fig.~\ref{fig:areas}) with the requirement that their magnetic footpoints had to be clearly visible 4 hours before the eruption and remain visible for at least 4 hours inside the dimming. This allowed us to compare the CBP evolution as the surrounding changes from a quiet Sun region to the dimming region. 

\begin{figure*}
    \centering
    \includegraphics[width=\linewidth]{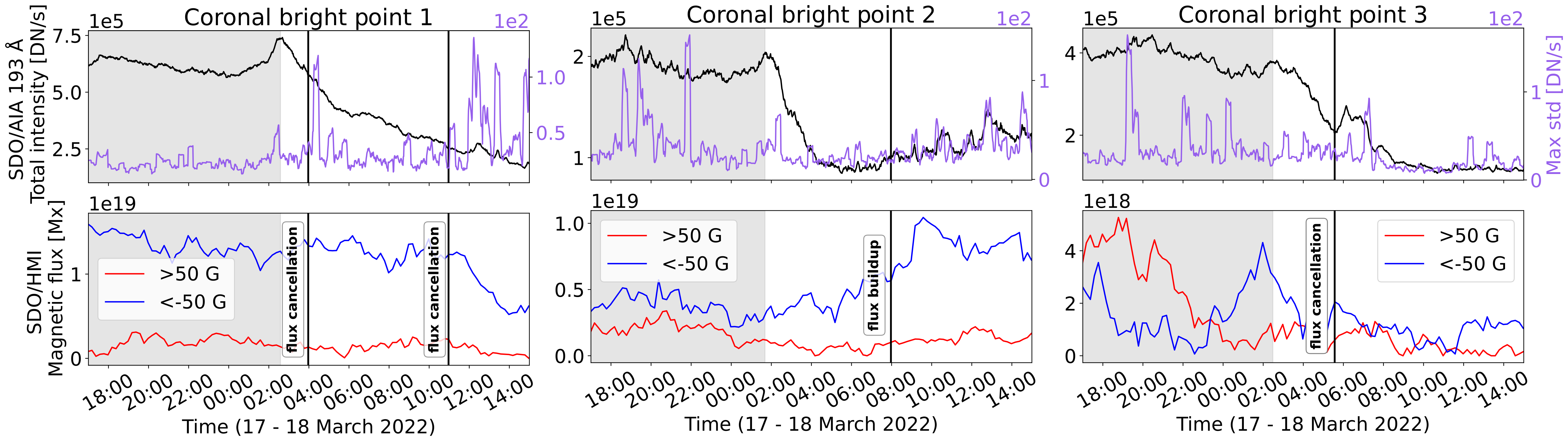}
    \caption{Temporal evolution of total intensity, standard deviation and magnetic flux at coronal bright point location from 17:00 UT on 17 March to 15:00 UT on 18 March 2022.
    Top panels: Total intensity (black curve) and maximum standard deviation computed over a 15-minute interval (purple curve), both derived from SDO/AIA 193\;{\AA} intensity data.  
Bottom panels: Absolute value of the total positive (red) and negative (blue) flux is depicted, where only pixels above/below $\pm 50$\;G are considered. The vertical black lines indicate times of flux cancellation or flux build-up. The interval before the onset of dimming is shaded in gray. The corresponding online movies display the temporal evolution of the coronal bright points in SDO/AIA 193 {\AA} images, overlaid with SDO/HMI LOS magnetic field contours at levels of $\pm25$\;G, $\pm50$\;G, $\pm100$\;G, $\pm150$\;G, and $\pm200$\;G (blue indicating negative polarity and red indicating positive polarity). The regions used for deriving the curves in this figure are indicated in the movie by a black-solid rectangle (for intensity and standard deviation) and a white-dashed rectangle (for magnetic flux), respectively. The associated movie is available online.}
    \label{fig:BPs_intensity_std_mag}
\end{figure*}

To quantify the dynamics of the three CBPs, in the SDO/AIA~193\; {\AA} images, we measured the total intensity (sum over all pixels) and the standard deviation of the intensity at the CBP locations (white rectangles in Fig~\ref{fig:areas}). The standard deviation was computed for each pixel over a 15-minute sliding window with 1-minute cadence. At each time step, we identify and select the value of the pixel with the maximum standard deviation. The temporal evolution of this is plotted in Fig.~\ref{fig:BPs_intensity_std_mag} (purple curve). High peaks correspond to transient brightenings at the CBP. We compared them to see whether there are more high peaks after the eruption than before.

We observed fewer standard deviation peaks in CBP 2 and 3 after the eruption. Coronal bright point~1 on the other hand shows only low activity before the eruption, but many standard deviation peaks after the eruption. These peaks coincide with flux cancellation, as illustrated in the lower panel of Fig.~\ref{fig:BPs_intensity_std_mag}. The flux cancellation at $\sim$10:30 UT is clearly visible as a pronounced decrease in the measured flux. In contrast, the flux cancellation at $\sim$04:00 UT is not as apparent in the plotted time series, as it does not involve the main footpoints. However, this cancellation event can be seen in the corresponding online movie at approximately (185\arcsec, -590\arcsec). Since the activity of CBP~1 is highly impacted by flux cancellation, it is not possible to determine with certainty whether some of this activity is influenced by the dimming. Therefore, we only concentrated on CBPS~2 and 3. We note that neither CBP~2 nor CBP~3 show any increase in standard deviation.

We derived the light curves by spatially integrating the SDO/AIA 193\;\AA\;intensity over the CBP regions (see Fig.~\ref{fig:areas}). They are plotted as black curves in Fig.~\ref{fig:BPs_intensity_std_mag}. When the dimming forms, the light curves exhibit a drop in intensity. Following this decrease, the intensity of CBP~2 rises again. This enhancement is not caused by small transient brightenings, but it is instead associated with an increase in size of the CBP and an overall intensity increase of the CBP loops. The expansion occurs because of a change in magnetic connectivity to a more distant positive-polarity footpoint. We surmise that the loops increase in overall intensity due to flux build-up, where the negative-polarity footpoint reaches field strengths below -200\;G at 07:58~UT. In CBP 3, we do see an additional intensity peak during the dimming phase. This peak occurs directly after the cancellation of one of the positive-polarity footpoints, which results in an eruption and connections to a new positive-polarity. After this eruption CBP~3 slowly faints. Overall, the CBP behaviours are therefore governed by the photospheric magnetic field behaviour.

\begin{figure}[!ht]
    \centering
    \includegraphics[width=1\linewidth, trim=0 0 0 0cm, clip=true]{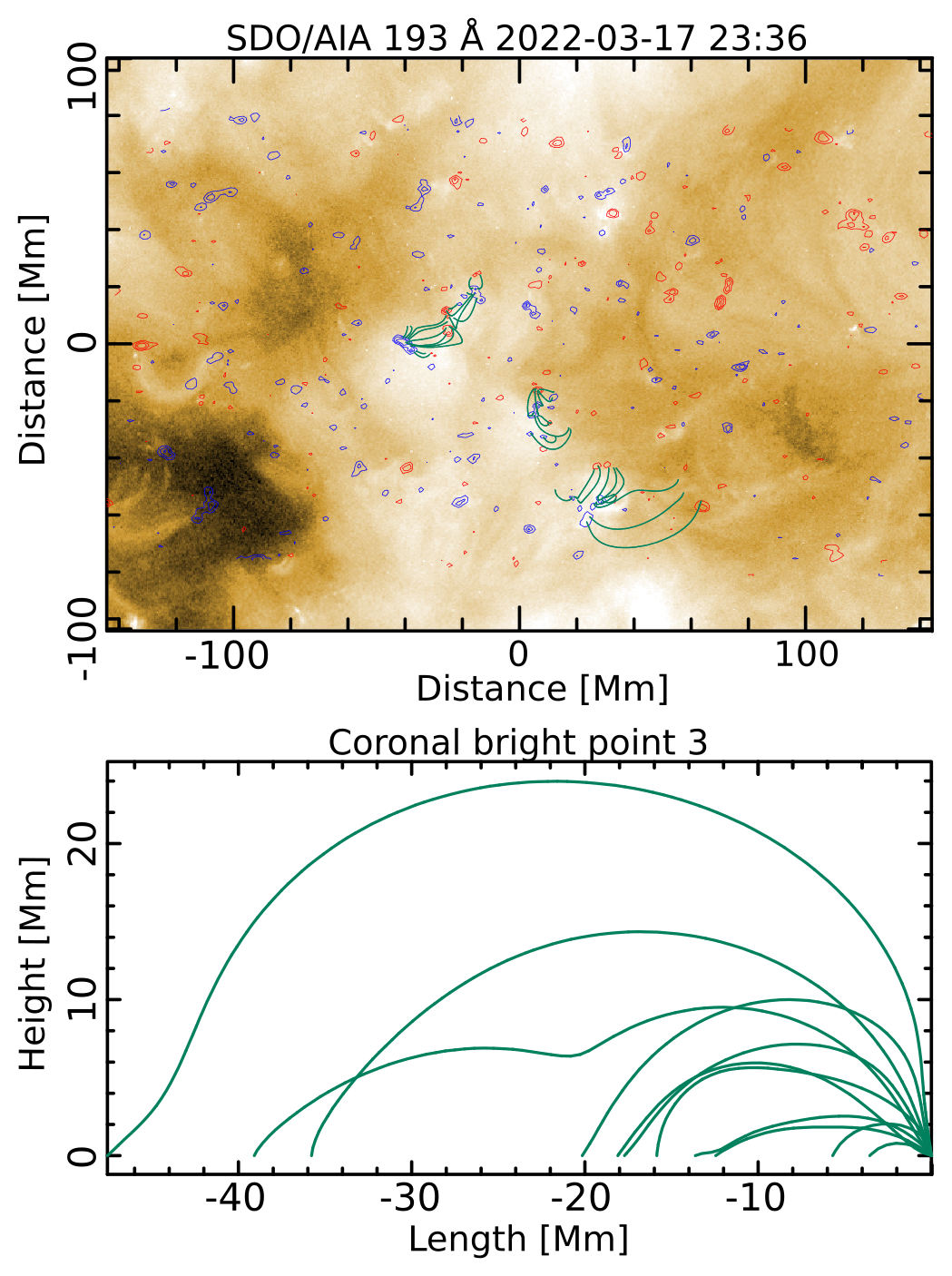}
    \caption{Magnetic field extrapolation at the coronal bright point locations. Top panel: SDO/AIA 193~{\AA} image with SDO/HMI magnetic field contours in red (positive) and blue (negative). The contours are at $\pm$20 G, $\pm$50 G and $\pm$100~G. The green lines show extrapolated magnetic field lines anchored at the flux concentrations where the three coronal bright points are located. The overlay between the magnetic model and AIA image takes into account the projection on the solar surface. The FOV corresponds to the dotted-black line in Fig.~\ref{fig:areas}. Bottom panel: Height of the field lines anchored at coronal bright point~3 as a function of the projected length. The height is measured orthogonal to the photospheric plane, while the length is measured along the projection of each field line onto that plane. For each field line, the start point is set to (0 Mm, 0 Mm).}
    \label{fig:BPs_mag_modelling}
\end{figure}

Using the magnetic field model described in Sect.~\ref{subsec:MagneticModelling}, we computed a set of field lines with starting footpoints at the photospheric flux concentrations where CBPs 1, 2, and 3 are located. Figure \ref{fig:BPs_mag_modelling} (top panel) shows the set of field lines in the vicinity of the three CBPs. From this set, we can estimate the height above the photosphere of the magnetic loops associated with each CBP. The height of the loops is plotted for CBP~3 at the bottom panel of Fig.~\ref{fig:BPs_mag_modelling}. The three longer field lines trace the magnetic connectivity away from the CBP. But within the core of the CBP itself, our extrapolation shows field lines reaching heights of $\sim 10$ Mm. Coronal bright point~2 has magnetic loops of $\sim 7$\;Mm height. 

\section{Discussion and conclusions}
In this work, we analysed EUV imaging data and spectroscopic data to compare a dimming region that formed within a quiet Sun location. We investigated whether the dimming event transforms the quiet Sun into an area that exhibits a typical coronal hole-like behaviour and how it might influence small-scale structures in the quiet Sun.

A quiet Sun filament eruption at $\sim$02:00 UT on 18 March 2022 caused a coronal dimming. The dimming had an initial fast growth phase for about 80 minutes with an area growth rate of $(2.3\pm 0.2) \cdot 10^6 \text{ km}^2\text{s}^{-1}$. \citet{hofmeisterFormationCoronalHole2025} found a similar growth rate of $1.7 \cdot 10^6 \text{ km}^2\text{s}^{-1}$ for the first 1.5 hours of a dimming also from a quiet Sun filament eruption. The growth rate is relatively slow compared to the growth-rate range of $[2.6, 76.7] \cdot 10^6 \text{ km}^2\text{s}^{-1}$ found by  \citet{dissauerStatisticsCoronalDimmings2019} in a data set of 62 dimmings. However, their values correspond to maximum growth rates and most of the events in their dataset are associated with active regions. Furthermore, they restricted their analysis to events occurring within $\pm 40^\circ$ of the central meridian. Because our event is located at $-40^\circ$ south, the derived value might underestimate the growth rate due to projection effects.

Within about 11 hours, the intensity measured with SDO/AIA 193\;{\AA} in the dimming region was reduced to the same level as the adjacent polar coronal hole. In the chromosphere, the dimming shows a $(9\pm 1)\%$ reduction in intensity relative to the pre-event quiet Sun. Similar decreases in the \ion{Mg}{II} k$_3$ intensities were reported in coronal holes \citep{upendranFormationSolarWind2022, kayshapQuietSunCoronalHole2018}. Furthermore, \citet{upendranFormationSolarWind2022} found that coronal holes have $\sim 2-12\%$ reduced intensities compared to the quiet Sun, with an increasing difference with stronger photospheric magnetic field $|B|$. 

We measured the flow characteristics of the dimming and compared it to the quiet Sun at the same location prior the filament eruption. Using the coronal Fe XII line measured with Hinode/EIS, we found a  $-2.9^{+0.2}_{-0.3}\,\mathrm{km\,s^{-1}}$ shift of Doppler velocity towards a greater upflow. This is comparable to previous studies for coronal holes, which reported that with the same line the Doppler velocity in a polar coronal hole is reduced, relatively to the surrounding quiet Sun, by $5.6\,\mathrm{km\,s^{-1}}$ \citep{tianNASCENTFASTSOLAR2010} and $\sim 3\,\mathrm{km\,s^{-1}}$\citep{kayshapDiagnosticsCoronalHole2015a}, respectively. This observation is consistent with the flow patterns studied for the same dataset by \citet{ngampoopunMergingCoronalDimming2023}. They found that approximately one hour after the eruption, the dimming region exhibits a pronounced blue shift with an additional secondary component in the blue wing, which can be interpreted as an indication of rapid plasma evacuation. After the merging of the dimming and the coronal hole, the plasma flows become more disordered, displaying a mixture of up- and downflows. Overall, a net upflow remains (without a secondary component), which reaches similar values to the southern coronal hole at $\sim13:30$~UT. In the upper chromosphere, we found comparable Doppler velocities between the dimming region and the quiet Sun. However, when only looking at upflows, the quiet Sun exhibits on average stronger upflows by $\sim 0.5\,\mathrm{km\,s}^{-1}$ for magnetic field strengths below $20\,\mathrm{G}$. \citet{upendranFormationSolarWind2022} found enhanced up- and downflows of $1-2$ km/s in coronal holes (latitudes and longitudes $<60^\circ$) compared to the quiet Sun. When considering the flow characteristics, the dimming region appears indistinguishable from a coronal hole in the corona, consistent with previous findings by \citet{ngampoopunMergingCoronalDimming2023}; however, in the upper chromosphere, the dimming does not exhibit the enhanced up- and downflows that have been reported for coronal holes. 

The composition of the dimming region started to resemble that of a coronal hole. We measured a lower FIP-bias in the dimming, compared to the quiet Sun at the same location prior the filament eruption. We find the FIP-bias to be closer to photospheric values in the dimming region with $1.4^{+0.1}_{-0.1}$. This value is still above the purely photospheric value of $\sim1$ expected in coronal holes \citep{bakerCoronalElementalAbundances2018, feldmanCoronalCompositionSolar1998}, which might be explained by the fact that we  used the \ion{Si}{X} 258.4~{\AA} / \ion{S}{X} 264.2~{\AA} intensity ratio as a FIP bias proxy, which non-linearly compresses the FIP bias (see Sect.~\ref{sec:eis}).

The activity of the CBPs, quantified through the temporal evolution of the 15-minute standard deviation, did not exhibit any systematic enhancement following the eruption. \citet{madjarskaEvidenceMagneticReconnection2004} and \citet{subramanianCoronalHoleBoundaries2010} found more transient brightenings around CBPs in coronal holes than in the quiet Sun. We would expect such transient brightenings to appear as peaks in the 15-minute standard deviation, but this behaviour was not observed. Therefore, we think that at least two out of the three CBPs are mostly unaffected by dimming. This result differs from the conclusions presented by \citet{ngampoopunMergingCoronalDimming2023}, who reported multiple brightenings and jets near the CBPs, many of which were associated with CBP 1. They associated these brightenings with the open magnetic configuration within the dimming region, which generally allows for more interchange reconnection and jet outflows. Coronal bright points are, in general, highly dynamic structures. Based on our comparison of the dynamics within the dimming region to those in the quiet Sun, we infer that we did not observe any enhancement in the number of brightenings beyond the typical variability exhibited by CBPs. Moreover, we excluded CBP~1 from our analysis because ongoing flux cancellation in the photospheric magnetic field produced numerous transient brightenings, making it difficult to relate any observed behaviour with the dimming event. Previous studies have also demonstrated that flux cancellation can give rise to brightenings such as jets \citep[e.g.][]{innesQuietSunExplosive2013, panesarMagneticFluxCancelation2016} and campfires \citep{panesarMagneticOriginSolar2021}. Given that only two CBPs were suitable for this study, additional observations are required to further validate the behaviour observed in this work.

The light curves at the CBP locations are mostly affected by the background intensity evolution. Coronal bright point~3 shows an additional peak due to an eruption of the CBP near the end of its lifetime. Such eruptions are commonly observed. \citet{mouEruptionsQuietSun2018a} found that 76\% of quiet-Sun CBPs produce at least one eruption during their lifetime, mostly during the late cancellation phase of their footpoints, after which the CBPs become smaller and fainter. Therefore, we think that this peak in intensity is part of normal CBP behaviour.

In the SDO/AIA 193~{\AA} images, we find an increased number of EUV brightenings within $10 \arcsec$ of the dimming border. This is consistent with the brightenings found along the coronal hole boundary by \citet{subramanianCoronalHoleBoundaries2010}. These brightenings could be a sign of reconnection taking place between the closed magnetic field lines of the quiet Sun and the open field of the dimming. We find this enhancement only in the hotter emission ($\sim 1.6$ MK), but not in the cooler SDO/AIA~171~\AA\ or Solar Orbiter HRI$_\text{EUV}$ 174~\AA\ ($\sim 1$ MK), suggesting that the reconnection-related brightenings predominantly occur at temperatures above 1 MK.  With SDO/AIA 193 {\AA}, we also find more brightenings inside the dimming contour compared to the pre-existing quiet Sun, but the increase is not significantly higher than the increase of brightenings outside the dimming. In SDO/AIA 171~{\AA}, we find a decrease in the number of brightenings inside the dimming which is in contrast to the findings by \citet{subramanianCoronalHoleBoundaries2010}. They found generally more brightenings in coronal holes compared to the quiet Sun. One possible explanation for this decrease is the more difficult detection due to an enhanced intensity in the dimming region, which we measured in the SDO/AIA 171\;{\AA} channel.  A comparable increase in the SDO/AIA {171~\AA} intensity was also reported by \citet{hofmeisterFormationCoronalHole2025} during a dimming event associated with a quiet-Sun filament eruption. They attributed this increase to the plasma cooling to approximately 1.1~MK. Within the core region of the dimming, the reported temperature decrease was consistent with adiabatic cooling.

In this paper, we show how a dimming event transforms the solar atmosphere and how it interacts with EUV brightenings. Coronal dimmings offer us a direct window into the mass loss linked to coronal mass ejections. Therefore, they can give us crucial insights into the acceleration of plasma in the solar wind. We find that the dimming transforms the quiet Sun into a coronal hole-like state only to a partial extent. We suggest that plasma is mainly removed from higher altitudes in the solar atmosphere, where we can observe spectroscopic properties and EUV intensity similar to those of a coronal hole. EUV brightenings inside the dimming, especially the bright points, do not change significantly compared to the behaviour in the previously quiet Sun. Therefore, we propose that the influence of the EUV wave and the resulting coronal dimming is confined to altitudes above a certain threshold. We refer to this characteristic altitude as the dimming height and propose that its lower limit is given by the magnetic loops associated with the bright points, which we estimate to be $\sim 10$ Mm above the surface. Along the dimming border (within 10\arcsec) in SDO/AIA $193$ \AA\ emission, we find an enhanced number of EUV brightenings. This result suggests that reconnection takes place at the border between the quiet Sun and the dimming region.  

\begin{acknowledgements} 
    SDO data are courtesy of NASA/SDO and the AIA, EVE, and HMI science teams.  
    Hinode is a Japanese mission developed and launched by ISAS/JAXA, with NAOJ as domestic partner and NASA and STFC (UK) as international partners. It is operated by these agencies in co-operation with ESA and NSC (Norway). IRIS is a NASA small explorer mission developed and operated by LMSAL with mission operations executed at NASA Ames Research center and major contributions to downlink communications funded by ESA and the Norwegian Space Centre. Solar Orbiter is a space mission of international collaboration between ESA and NASA, operated by ESA. The EUI instrument was built by CSL, IAS, MPS, MSSL/UCL, PMOD/WRC, ROB, LCF/IO with funding from the Belgian Federal Science Policy Office (BELSPO/PRODEX PEA 4000112292 and 4000134088); the Centre National d’Etudes Spatiales (CNES); the UK Space Agency (UKSA); the Bundesministerium für Wirtschaft und Energie (BMWi) through the Deutsches Zentrum für Luft- und Raumfahrt (DLR); and the Swiss Space Office (SSO). This research used version 7.0.0 \citep{stuart_j_mumford_2025_15691296} of the SunPy open source software package \citep{sunpy_community2020}. CHIANTI is a collaborative project involving George Mason University, the University of Michigan (USA), University of Cambridge (UK) and NASA Goddard Space Flight Center (USA). This work used data provided by the MEDOC data and operations centre (CNES / CNRS / Univ. Paris-Saclay)\footnote{\url{http://medoc.ias.u-psud.fr/}}, and by JSOC \footnote{\url{http://jsoc.stanford.edu/}}.
    The work of D.H.B. was performed under contract to the Naval Research Laboratory and was funded by the NASA Hinode program.
    ACS received support from the Heliophysics Division of NASA’s Science Mission Directorate through the Heliophysics Supporting Research (HSR) Program,  and from the NASA/MSFC Hinode Project, and he also benefited from discussions at the International Space Science Institute (ISSI-BJ ID 24-604) team meeting on ``Small-scale eruptions in the Sun''. 
    L.P.C. gratefully acknowledges funding by the European Union (ERC, ORIGIN, 101039844).
\end{acknowledgements}

\bibliographystyle{bibtex/aa}
\bibliography{bibtex/bib}
\begin{appendix}
\onecolumn

\clearpage

\end{appendix}
\end{document}